\documentstyle[aps,epsfig]{revtex}

\begin{document}

\begin{center}
{\Large \bf
Report from NA49
}
\end{center}

\vspace{0.5cm}
\noindent
M. Ga\'zdzicki$^{9,12,}$\footnote{invited talk presented at Quark Matter 2004}, 
C.~Alt$^{9}$, T.~Anticic$^{21}$, B.~Baatar$^{8}$,D.~Barna$^{4}$,
J.~Bartke$^{6}$, 
L.~Betev$^{9,10}$, H.~Bia{\l}\-kowska$^{19}$, A.~Billmeier$^{9}$,
C.~Blume$^{9}$,  B.~Boimska$^{19}$
% ...
\begin{center}
(NA49 Collaboration)
\end{center}

\vspace{0.2cm}
\noindent
% ...
$^{4}$KFKI Research Institute for Particle and Nuclear Physics, Budapest, Hungary.\\
% ...
$^{6}$Institute of Nuclear Physics, Cracow, Poland.\\
% ...
$^{8}$Joint Institute for Nuclear Research, Dubna, Russia.\\
$^{9}$Fachbereich Physik der Universit\"{a}t, Frankfurt, Germany.\\
$^{10}$CERN, Geneva, Switzerland.\\
% ...
$^{12}$Institute of Physics \'Swi{\,e}tokrzyska Academy, Kielce, Poland.\\
% ...
$^{19}$Institute for Nuclear Studies, Warsaw, Poland.\\
% ...
$^{21}$Rudjer Boskovic Institute, Zagreb, Croatia.\\

\vspace{0.5cm}
\begin{abstract}
\noindent
The most recent data of NA49 on hadron production in nuclear collisions
at CERN SPS energies are presented.
Anomalies in the energy dependence of pion and kaon production in 
central Pb+Pb collisions are observed. They suggest that the
onset of deconfinement is located at about 30 $A$GeV.  
Large multiplicity and transverse momentum fluctuations
are measured for collisions of intermediate mass systems at 158 $A$GeV.
The need for a new experimental programme at the CERN SPS is 
underlined. 
\end{abstract}

\newpage

\vspace{0.2cm}
\noindent
{\bf 1. Introduction}\\ 
Since the first Pb-run at the CERN SPS (at 158 $A$GeV) in 1994
the NA49 experiment \cite{Afanasev:1999iu} collected a large set of
data on hadron
production in nuclear collisions.
The data taking period ended in 2002 with Pb-runs 
at 20 and 30 $A$GeV. 
Several research programmes were undertaken by the NA49
Collaboration.
Selected results of two of them, the energy scan programme and
the system size dependence programme, are reported here.
Other recent results of NA49 on the energy dependence of particle
ratio fluctuations~\cite{christof}, 
(multi)strange hyperon production~\cite{christine} and two pion
correlations~\cite{stefan} were also presented during this conference.
Finally our evidence for pentaquark candidates in p+p interactions
at 158 $A$GeV was discussed~\cite{kreso}. 

The energy scan programme was motivated by the hypothesis 
\cite{Gazdzicki:1998vd} that the onset of 
the deconfinement phase transition is located 
between the top SPS and AGS energies.
Within this project data on hadron production in central 
Pb+Pb collisions at 20, 30, 40, 80 and 158 $A$GeV were recorded.
In this report we show  the first  results obtained at 20 $A$GeV
which extend the  previously measured hadron systematics 
\cite{Afanasiev:2002mx,Alt:2003rn} to the full SPS energy range
from 20 to 158 $A$GeV.

The aim of the system size dependence programme is to study
how the properties of strongly interacting matter
change with its volume.
Data on p+p, C+C, Si+Si and minimum bias Pb+Pb
collisions at 158 $A$GeV and 40 $A$GeV were collected.
Here we show new results on system size dependence of electric
charge correlations and multiplicity fluctuations at 158 $A$GeV.

\vspace{0.2cm}
\noindent
{\bf 2. Energy dependence of hadron production in central Pb+Pb collisions} 
\vspace*{-0.4cm}
\begin{figure}
\begin{center}
\mbox{ \epsfig{file=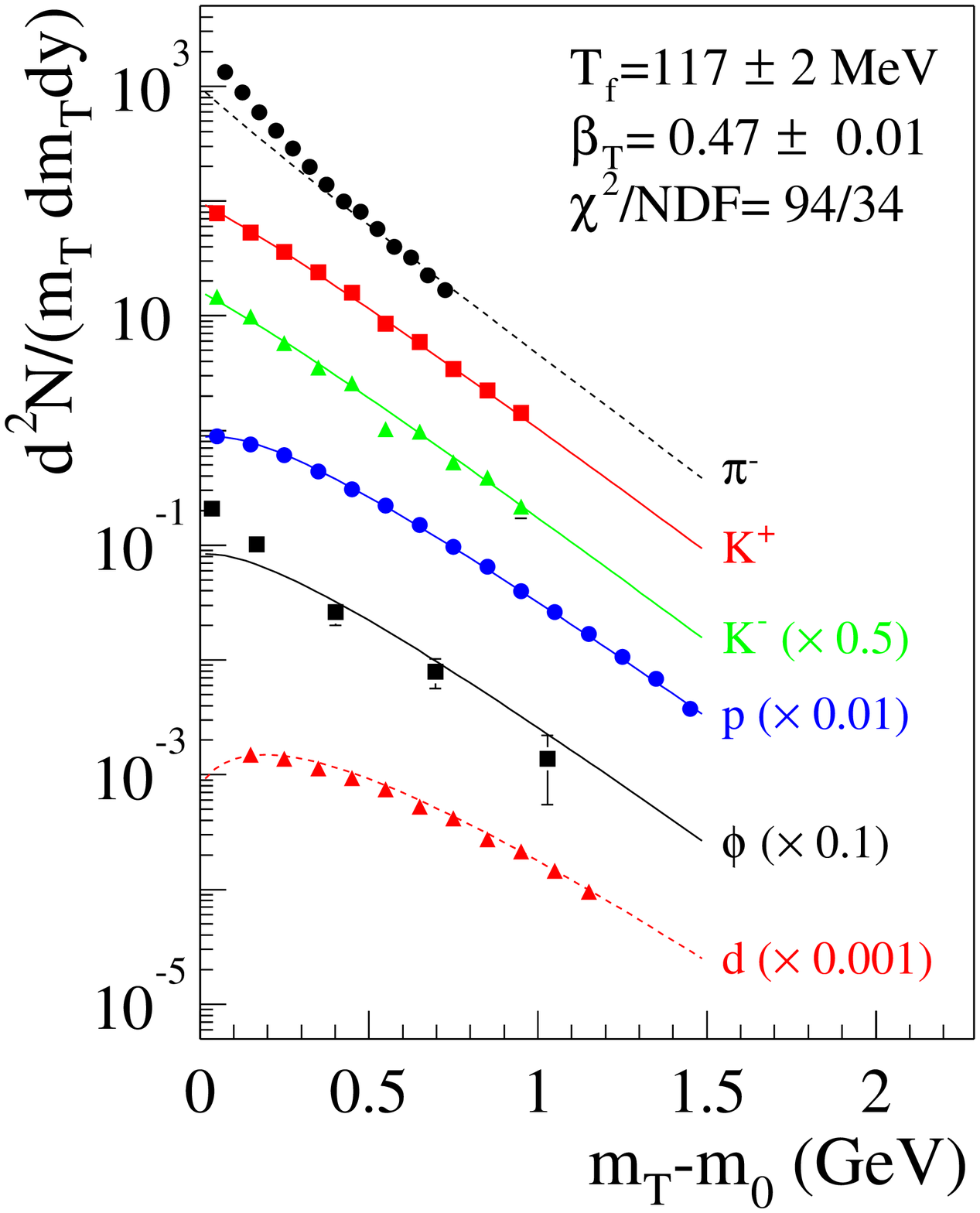,width=60mm} 
       \epsfig{file=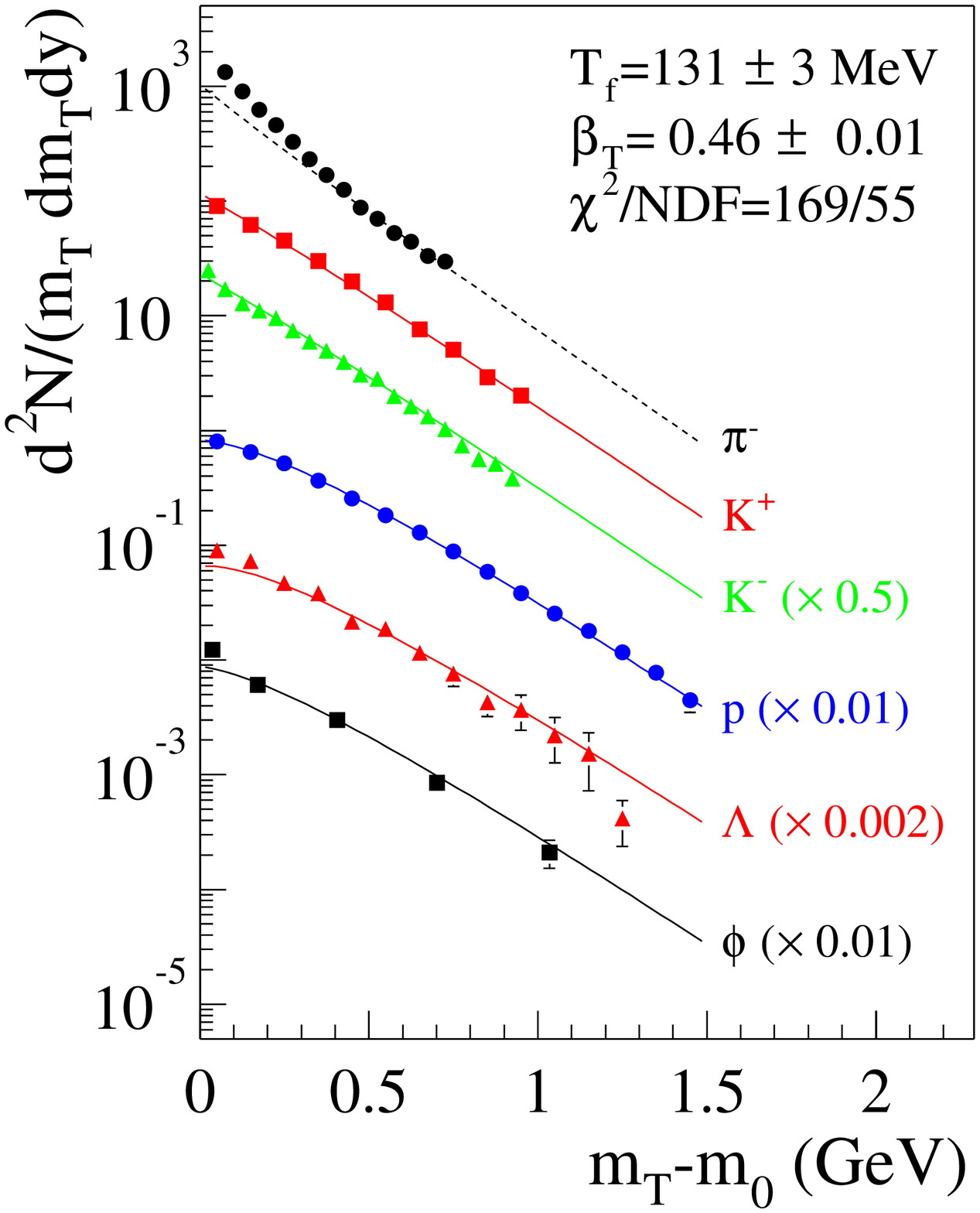,width=60mm} }
\end{center}
\caption{
Transverse mass spectra of hadrons produced in
central Pb+Pb collisions at 20 (left) and 30 (right)  $A$GeV.
The solid lines indicate fits of a blast wave parametrization
{\protect\cite{Schnedermann:gc}}.
}
\label{bw}
\end{figure}

The transverse mass, $m_T$, spectra of hadrons measured in central
(7.2\%) Pb+Pb collisions at 20 and 30 $A$GeV are presented in Fig.~\ref{bw}.
The data follow approximately the pattern expected within hydrodynamical
approaches.  
As an illustration the fits of the blast wave parametrization 
\cite{Schnedermann:gc} are shown.
The thermal freeze-out temperature and the collective transverse
velocity resulting from the fits are about 120 MeV  and 0.5$c$, 
respectively.
Similar values of  these parameters were obtained  at the higher SPS 
energies.

\begin{figure}
\begin{center}
\mbox{ \epsfig{file=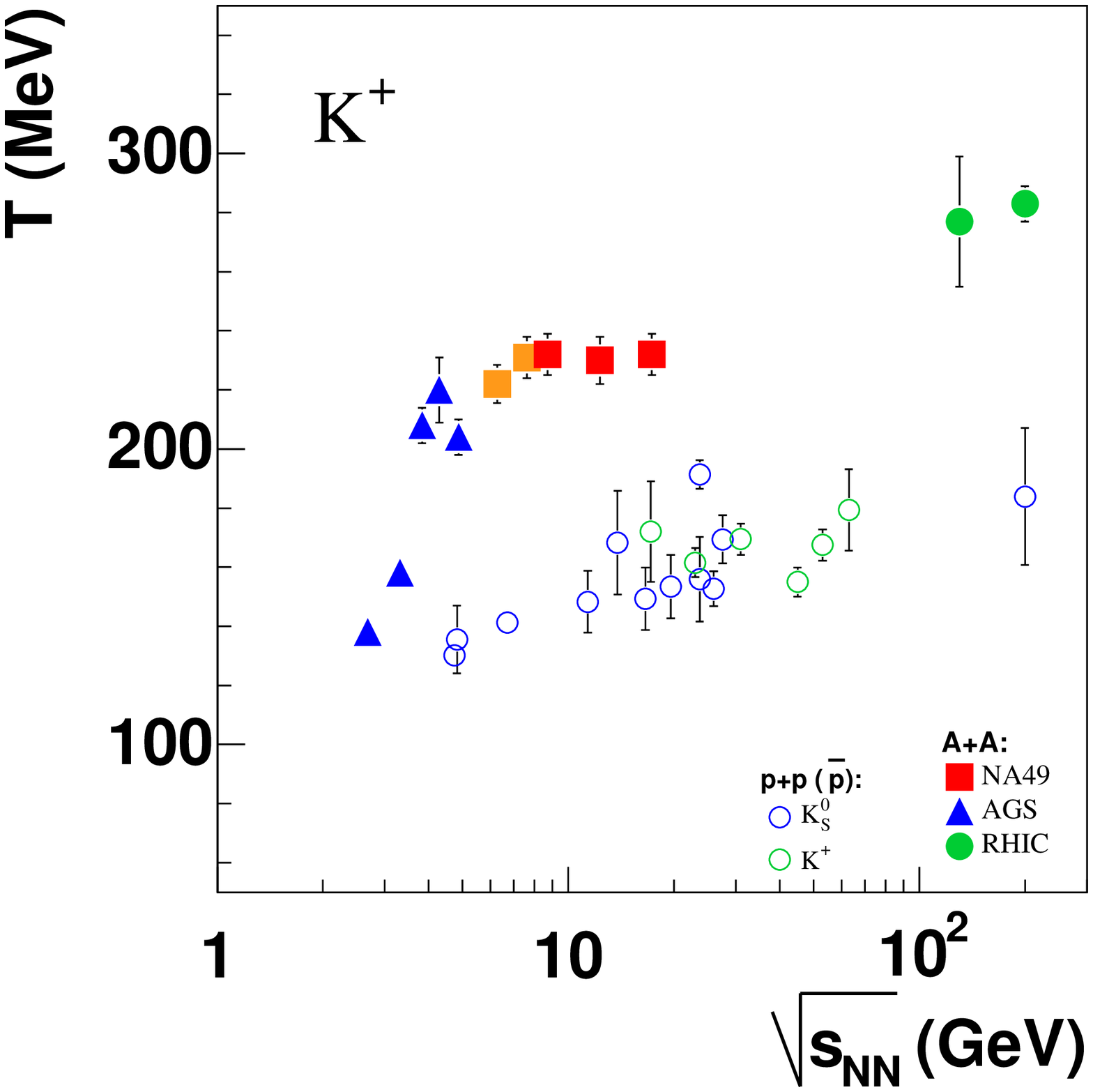,width=60mm} 
       \epsfig{file=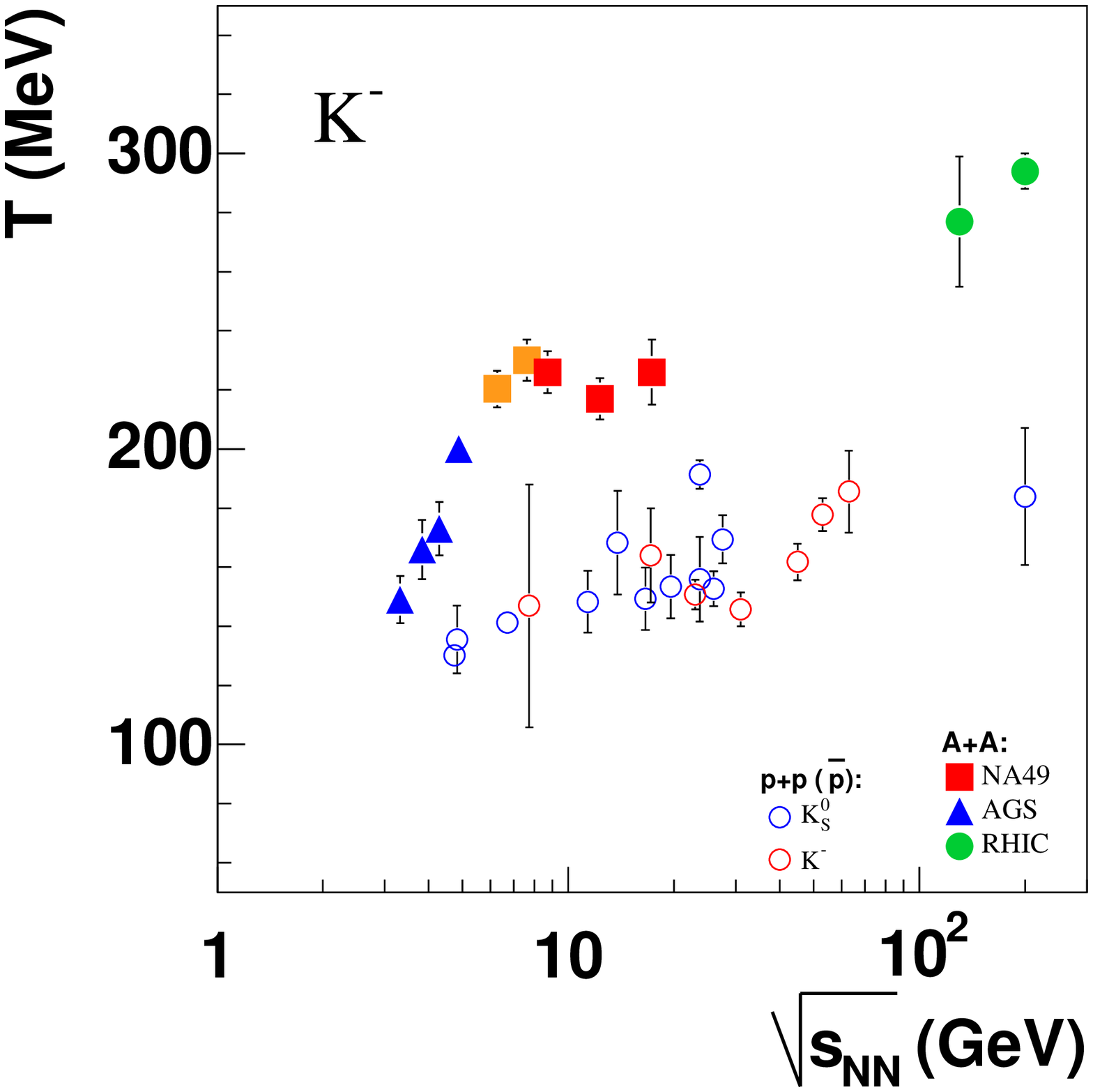,width=60mm} }
\end{center}
\caption{
Energy dependence of the inverse slope parameter, $T$,
of the transverse mass spectra of $K^+$ (left) and $K^-$
(right) produced in central Pb+Pb (Au+Au) collisions
(solid symbols) and p+p interactions (open symbols).
$\sqrt{s_{NN}}$ is the c.m.s. energy per nucleon-nucleon pair.
}
\label{slopes}
\end{figure}

The shape of the $m_T$-spectra of kaons appears to be in a good
approximation exponential, $1/m_T dn/dm_T \sim exp(-m_T/T)$,
in the full studied energy range.
Thus the shape of the spectra is well represented by their
inverse slope parameter $T$.
Its dependence on the collision energy is shown in 
Fig.~\ref{slopes} for central Pb+Pb~(Au+Au) collisions 
\cite{Gorenstein:2003cu} and p+p($\overline{\rm p}$)
interactions \cite{Kliemant:2003sa}.
For heavy ion collisions a steep rise at AGS energies is
followed by a plateau at SPS energies. At RHIC higher values
are observed.
The beginning of the plateau in
this 
{\bf step}-like dependence \cite{Gorenstein:2003cu}
is located 
at about 30 $A$GeV.

\begin{figure}
\begin{center}
\mbox{ \epsfig{file=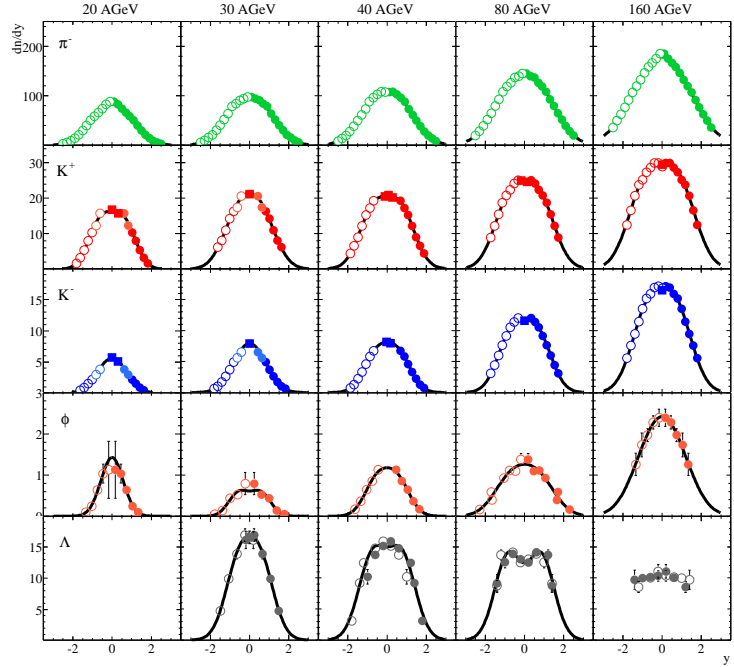,width=100mm} 
        }
\end{center}
\caption{
The rapidity spectra of hadrons produced in central (7\% at 
20-80 $A$GeV, 5\% ($\pi^-, K^+, K^-$) and 10\% ($\phi, \Lambda$)
at 158 $A$GeV) at SPS energies.
The closed symbols indicate measured points, open points are
reflected with respect to midrapidity.
The solid lines indicate parameterizations of the data used for the
extrapolation of the yield to full phase space.
}
\label{rap}
\end{figure}

\newpage

Rapidity $y$
($y$ is the rapidity of a particle in the collision
center-of-mass system) 
distributions of selected hadrons measured in central
Pb+Pb collisions at 20-158 $A$GeV are plotted in Fig.~\ref{rap}.
The large acceptance of the NA49 spectrometer and high
resolution in particle identification \cite{Afanasev:1999iu} 
allow for  reliable
measurements of  total mean multiplicities (denoted as 
$\langle ... \rangle$) of various hadronic species. 

\begin{figure}

\mbox{ \epsfig{file=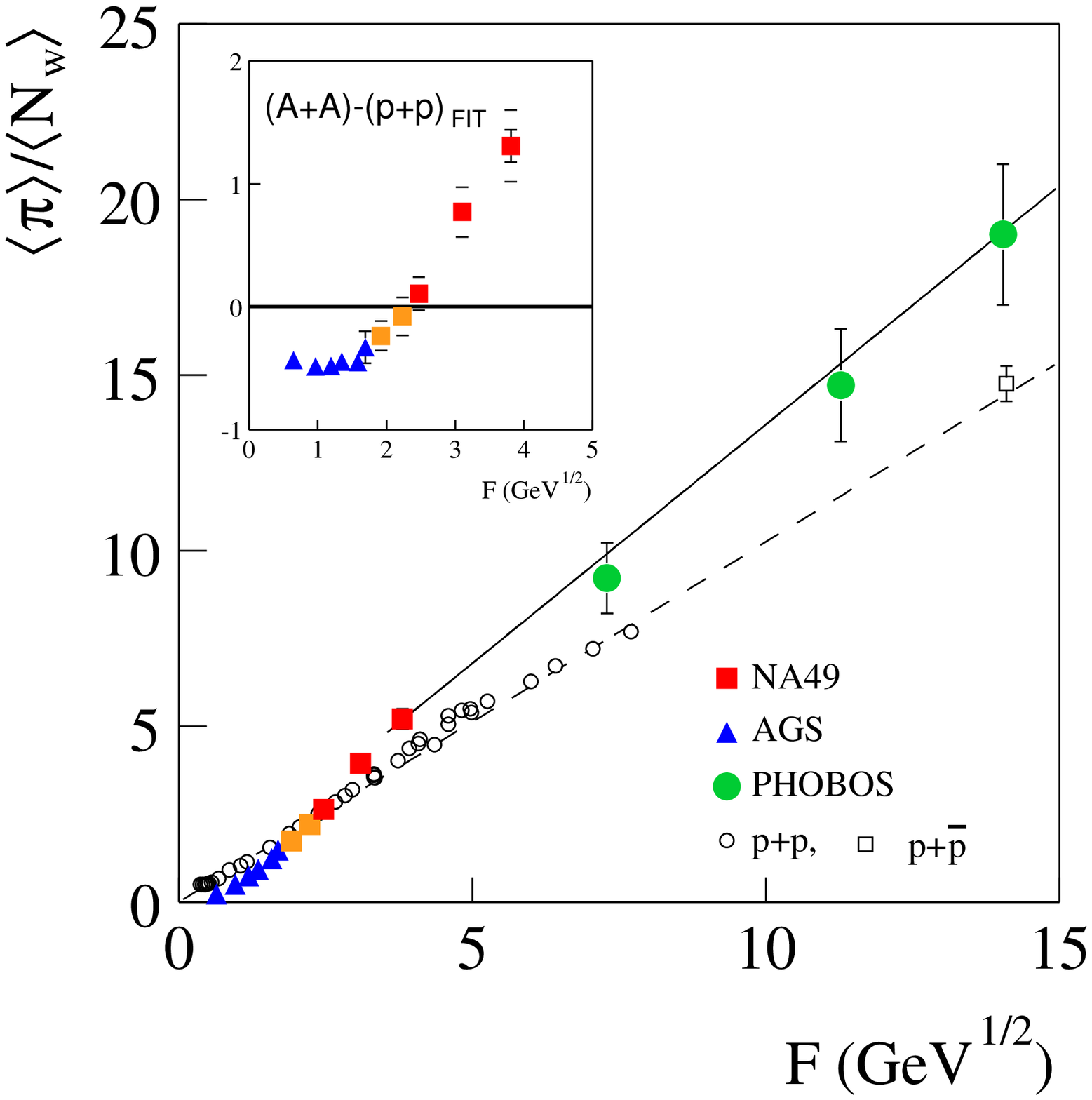,width=70mm} 
       \epsfig{file=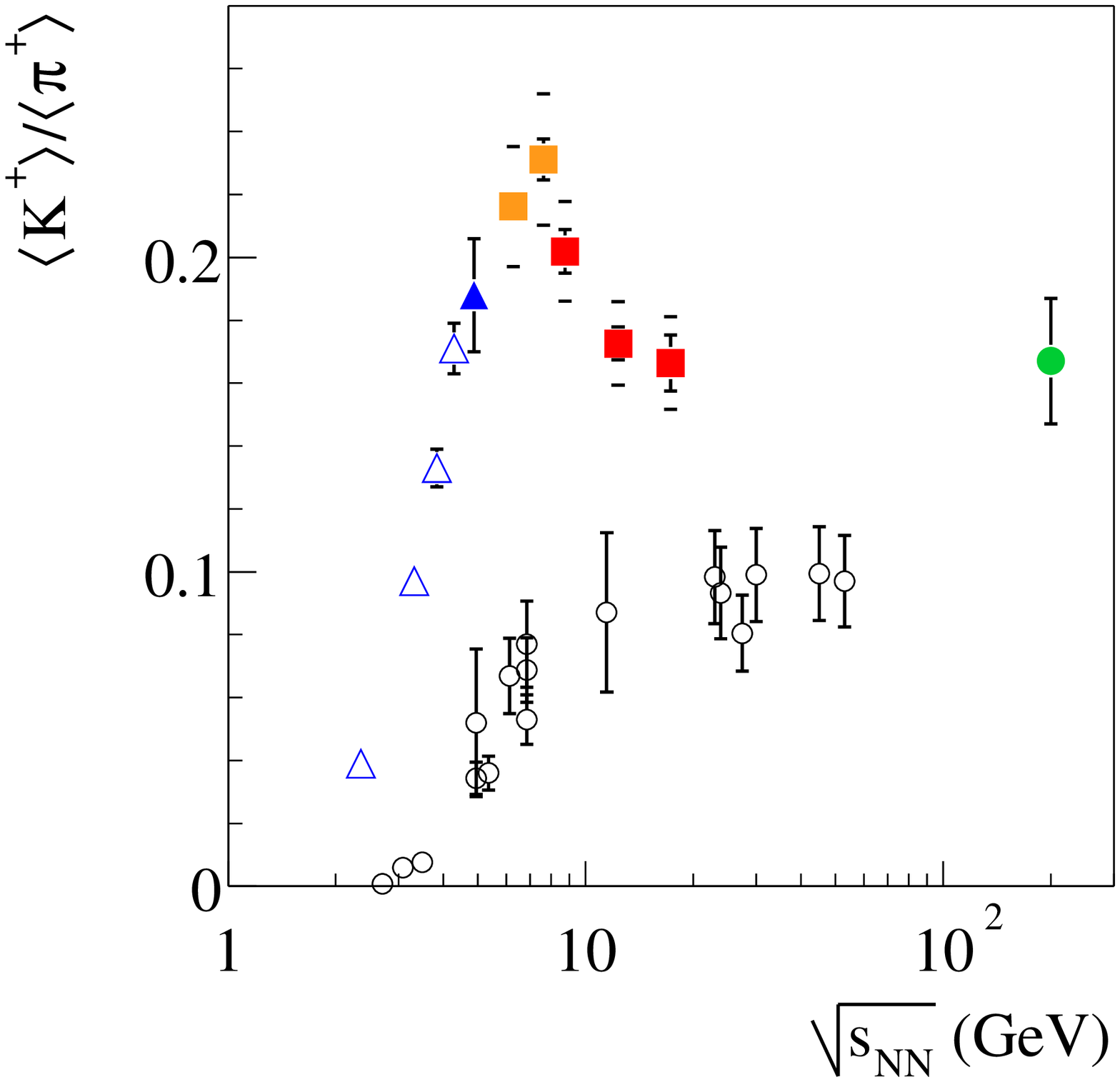,width=70mm} }

\caption{
Left: 
The dependence of total pion multiplicity per
wounded nucleon on Fermi's energy measure $F$
($F \equiv (\sqrt{s_{NN}}
- 2 m_N)^{3/4}/\sqrt{s_{NN}}^{1/4}$,
where $\sqrt{s_{NN}}$ is the c.m.s. energy
per nucleon--nucleon pair and $m_N$ the rest mass of the nucleon)
for central Pb+Pb (Au+Au)  collisions (closed symbols)
and inelastic p+p($\overline{\rm p}$) interactions
(open symbols).
Right:
The dependence of the $\langle K^+ \rangle/\langle \pi^+ \rangle$
ratio on the collision energy for
central Pb+Pb (Au+Au) collisions (closed symbols)
and inelastic p+p interactions
(open symbols).
}
\label{pika}
\end{figure}

The dependence of mean multiplicity of pions 
($\langle \pi \rangle = 
   1.5 \cdot (\langle \pi^- \rangle + \langle \pi^+ \rangle)$ 
per mean number of wounded nucleons, $\langle N_W \rangle$,
on collision energy is shown in Fig.~\ref{pika} (left)
for central Pb+Pb(Au+Au) collisions and 
p+p($\overline{\rm p}$) interactions.
A {\bf kink}-like change from pion suppression to  pion
enhancement in central Pb+Pb (Au+Au) collisions is observed 
at about 30 $A$GeV
in contrast to a linear increase seen for 
p+p($\overline{\rm p}$) interactions in the full energy range.

The $\langle K^+ \rangle/\langle \pi^+ \rangle$ ratio is plotted
as a function of collision energy in Fig.~\ref{pika} (right).
A {\bf horn}-like structure is observed for central 
Pb+Pb~(Au+Au) collisions. 
The maximum of the horn is located at about 30 $A$GeV.

The anomalies observed in the energy dependence of hadron production
(the {\bf horn}, the {\bf kink} and the {\bf step}) 
at the low SPS energies
are consistent with the predictions  for the onset
of the deconfinement phase transition \cite{Gazdzicki:1998vd}.
These anomalies can not be reproduced with current string-hadronic
(see e.g. Ref.~\cite{Bratkovskaya:2004kv}) 
and hadron gas models (see e.g. Ref.~\cite{Cleymans:1999st}).

\vspace{0.2cm}
\noindent
{\bf 2. System size  dependence of hadron production at 158 $A$GeV }

\begin{figure}
\begin{center}
\mbox{ \epsfig{file=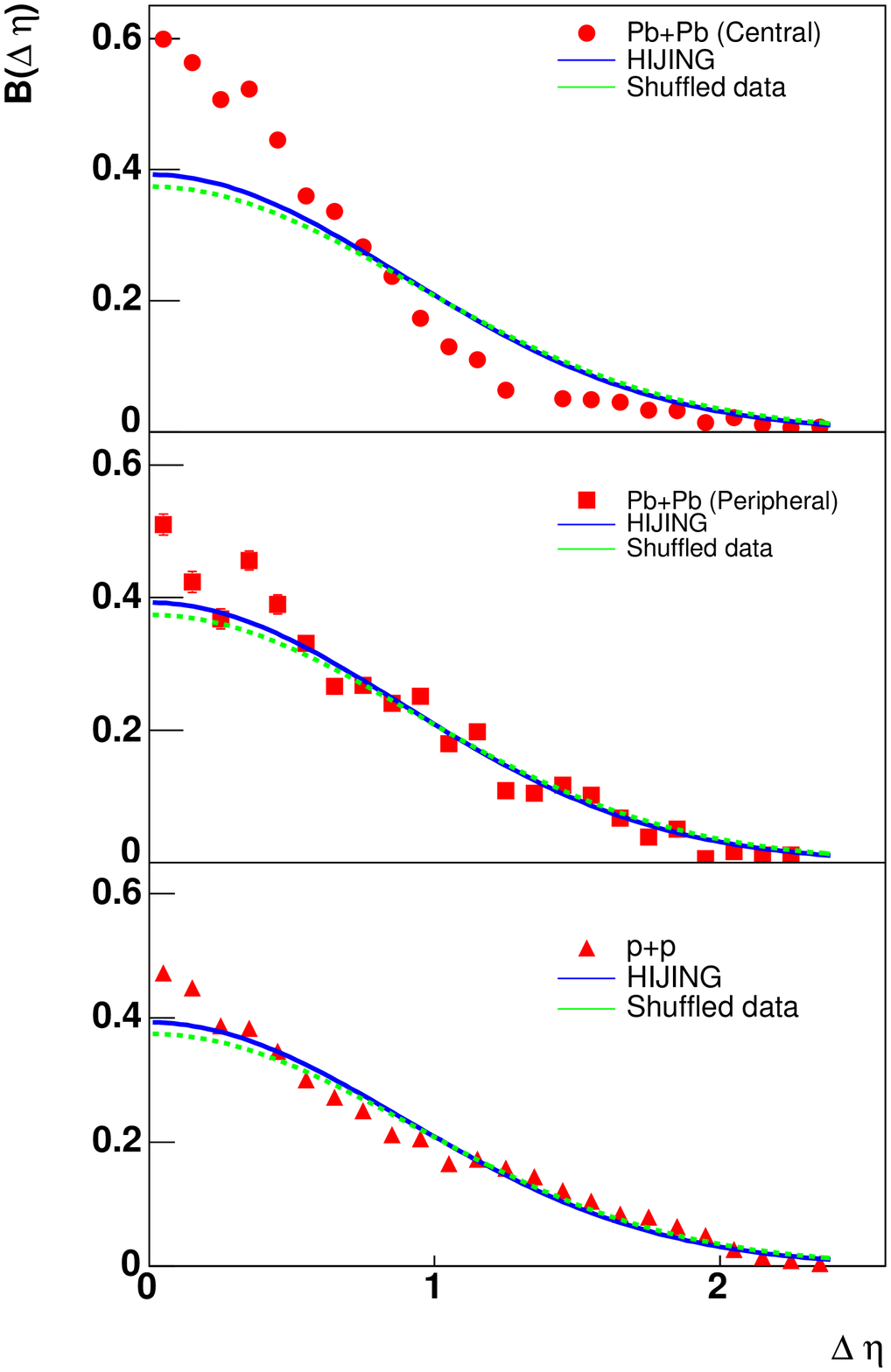,width=70mm} 
       \epsfig{file=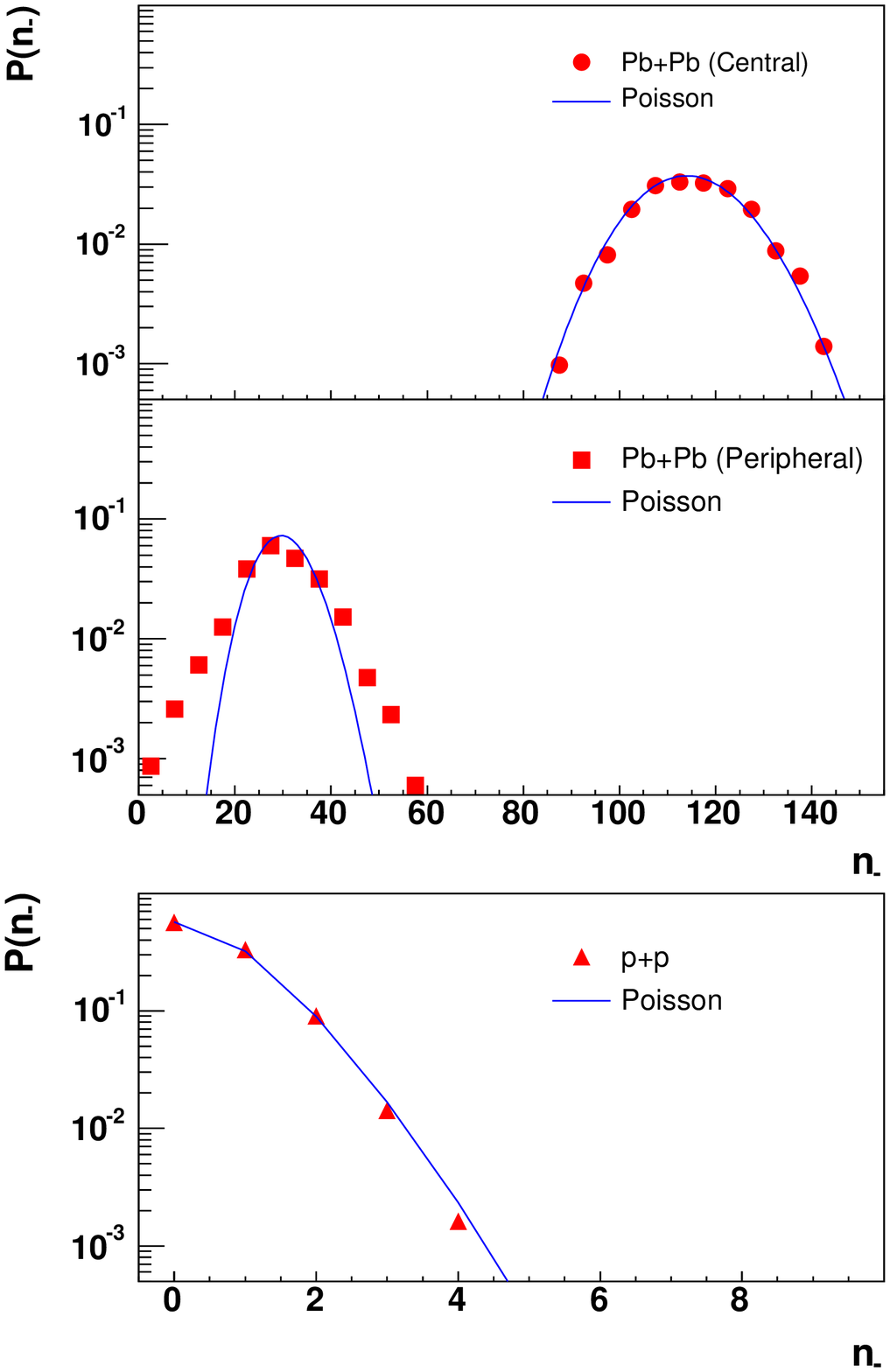,width=90mm} }
\end{center}       

\caption{
Left: The dependence of the balance function,
$B(\Delta \eta)$,
on pseudo-rapidity interval, $\Delta \eta$, for
selected reactions at 158 $A$GeV.
Right: The multiplicity distributions of negatively charged
particles, $P(n_-)$ for selected reactions at 158 $A$GeV. 
}
\label{mult}
\end{figure}

The correlations between positively and negatively charged
hadrons were studied in p+p, C+C, Si+Si and Pb+Pb collisions
at 158 $A$GeV \cite{panos}
in terms of the balance function  \cite{Bass:2000az}:
\begin{equation}
B(\Delta \eta) =
0.5 ( (N_{+-}(\Delta \eta) + N_{--}(\Delta \eta) )
/N_{-} + (N_{-+}(\Delta \eta) + N_{++}(\Delta \eta) )/N_{+} ),
\end{equation}
where $N_{ab}(\Delta \eta)$ is the number of pairs 
of particles of charges
$a$ and $b$ separated by a pseudo-rapidity interval $\Delta \eta$
and $N_-$, $N_+$ are total numbers of negatively and positively
charged particles used in the analysis, respectively.
Hadrons  within the acceptance limits,
$0.005 < p_T < 1.5$ GeV/c and $2.6 < \eta < 5.0$
($p_T$ and $\eta$ are transverse momentum and pseudo-rapidity in
the laboratory system, respectively),
which passed in addition the NA49 
acceptance filter were used for the analysis.
About 30\% of all produced charged hadrons are accepted. 
As an example, the balance functions for central and peripheral
Pb+Pb collisions and p+p interactions are 
shown in Fig.~\ref{mult} (left).
The distribution measured for central Pb+Pb collisions is
significantly narrower than the spectra obtained for
the string-hadronic model Hijing \cite{Gyulassy:ew} 
and ``shuffled'' \cite{Adams:2003kg}  events.
``Shuffled'' events are constructed from the real events by
random redistribution of particle pseudo-rapidities within one event.

\begin{figure}
\begin{center}
\mbox{ \epsfig{file=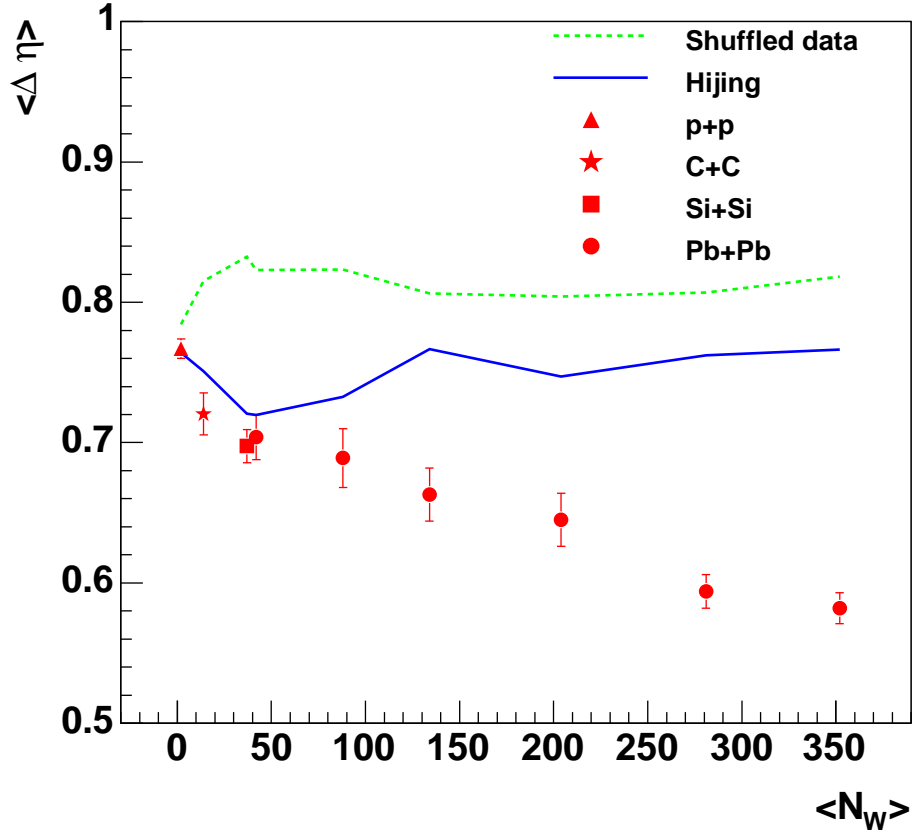,angle=270,width=120mm} }
\end{center}
\vspace*{0.5cm}
\caption{
The dependence of the width of the balance function,
$\langle \Delta \eta \rangle$,  on the mean number of
wounded nucleons, $\langle N_W \rangle$,
for various reactions measured at 158 $A$GeV.
The experimental data are compared with the corresponding results
for  Hijing and ``shuffled'' events.
}
\label{bf_width}
\end{figure}

The width of the balance function in pseudo-rapidity 
calculated as 
\begin{equation}
\langle \Delta \eta \rangle =
\int^{2.4}_{0.1} \Delta \eta B(\Delta \eta) d\Delta \eta/
\int^{2.4}_{0.1}  B(\Delta \eta) d\Delta \eta 
\end{equation}
is plotted in Fig.~\ref{bf_width} 
as a function of $\langle N_W \rangle$ for all
studied reactions at 158 $A$GeV.
A monotonic decrease of $\langle \Delta \eta \rangle$
 with increasing $\langle N_W \rangle$
is observed. 
The width of the balance function obtained for Hijing  and
``shuffled''  events is approximately independent of 
 $\langle N_W \rangle$, see Fig.~\ref{bf_width}.
The reduction of  $\langle \Delta \eta \rangle$ 
 expressed by the ratio 
$(\langle \Delta \eta \rangle(data) -
\langle \Delta \eta \rangle(shuffled))/
\langle \Delta \eta \rangle(shuffled)) $
amounts to about 30\%  for central Pb+Pb collisions at 158 $A$GeV.
A similar reduction was measured in central Au+Au collisions
at RHIC energies \cite{Adams:2003kg}.
Originally a narrowing of the balance function was predicted
as a signal of late hadronization \cite{Bass:2000az}.
Recently  
the observed effect at RHIC energies is partly
explained
within the models which
assume a
transverse flow of freezing-out
matter and either quark coalescence \cite{Bialas:2003bb} or
decay of resonances \cite{Florkowski:2004qp}.

The multiplicity distributions of negatively charged hadrons
produced in p+p interactions as well as peripheral and central
Pb+Pb collisions at 158 $A$GeV are shown in 
Fig.~\ref{mult} (right).
Only hadrons in the forward hemisphere 
($1.1 < y < 2.6$)
which passed the NA49
acceptance filter \cite{Anticic:2003fd} were used for the analysis.
About 15\%  of all negatively charged hadrons are accepted.
For Pb+Pb collisions the fluctuations in the number of
projectile spectators, and thus in the number 
of projectile participants, were reduced by selecting 
narrow bins in the forward energy measured by the NA49 Veto
calorimeter \cite{Afanasev:1999iu}.
The distribution for p+p interactions and central Pb+Pb
collisions  are similar to a Poisson distribution (solid lines
in Fig.~\ref{mult} (right)).
For peripheral Pb+Pb collisions the measured distribution
is significantly broader than  Poisson.

\begin{figure}

\begin{center}
\mbox{ \epsfig{file=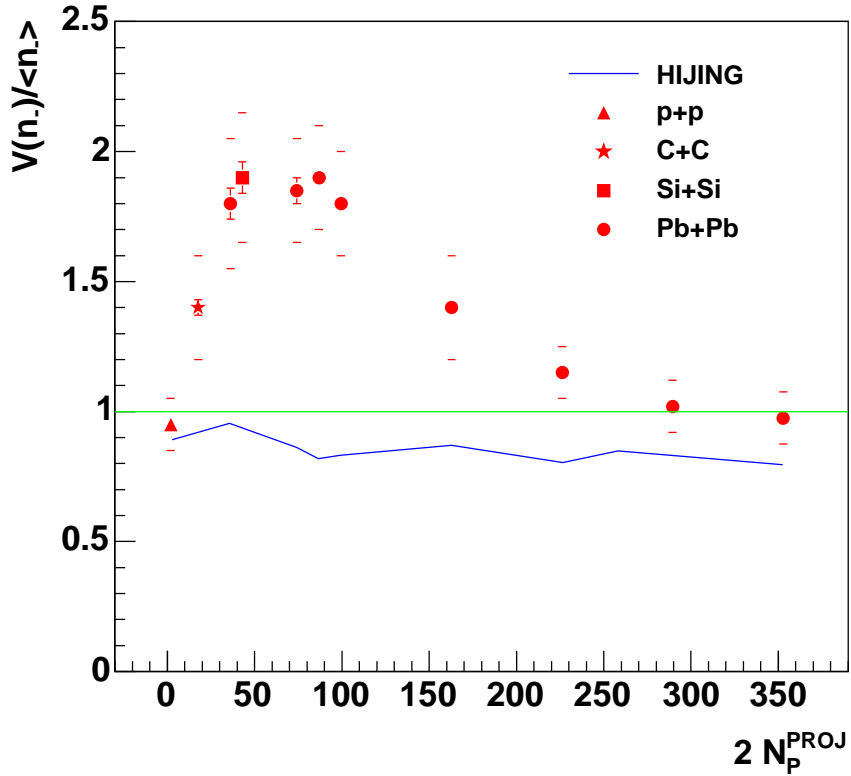,width=130mm} 
        }
\end{center}

\caption{
The dependence of the ratio of variance to mean  for the 
multiplicity distributions of negatively charged hadrons
on the number of projectile participants, $N_P^{PROJ}$,
measured in various reactions at 158 $A$GeV.
Results obtained within the string-hadronic model 
Hijing  are shown by the
solid line. 
}
\label{varn}
\end{figure}

The ratio of the variance ($V(n_-) = 
\langle n_-^2 \rangle - \langle n_- \rangle^2$)  to the mean
($\langle n_- \rangle$) 
of the multiplicity distributions of negatively charged particles 
is plotted in Fig.~\ref{varn} 
as a function of the
number of projectile participants, $N_P^{PROJ}$.
The ratio was corrected for the fluctuations in  $N_P^{PROJ}$
because of the finite width of the  bins in the energy measured by the
Veto calorimeter and its resolution \cite{maciek}.
The dependence of the ratio  $V(n_-)/\langle n_- \rangle$ on
$N_P^{PROJ}$ exhibits a maximum at about 
$2 N_P^{PROJ} \approx N_W \approx 70$.
Note, that the transverse momentum fluctuations expressed by the
$\Phi_{p_T}$ measure show a similar behaviour \cite{kasia,Anticic:2003fd}.
The observed large fluctuations for the intermediate mass
systems were not predicted. Their origin is unclear.

\vspace{0.2cm}
\noindent
{\bf 4. Summary and the new programme at the CERN SPS } 

\begin{figure}
\vspace*{-3cm}
\begin{center}
\mbox{ \epsfig{file=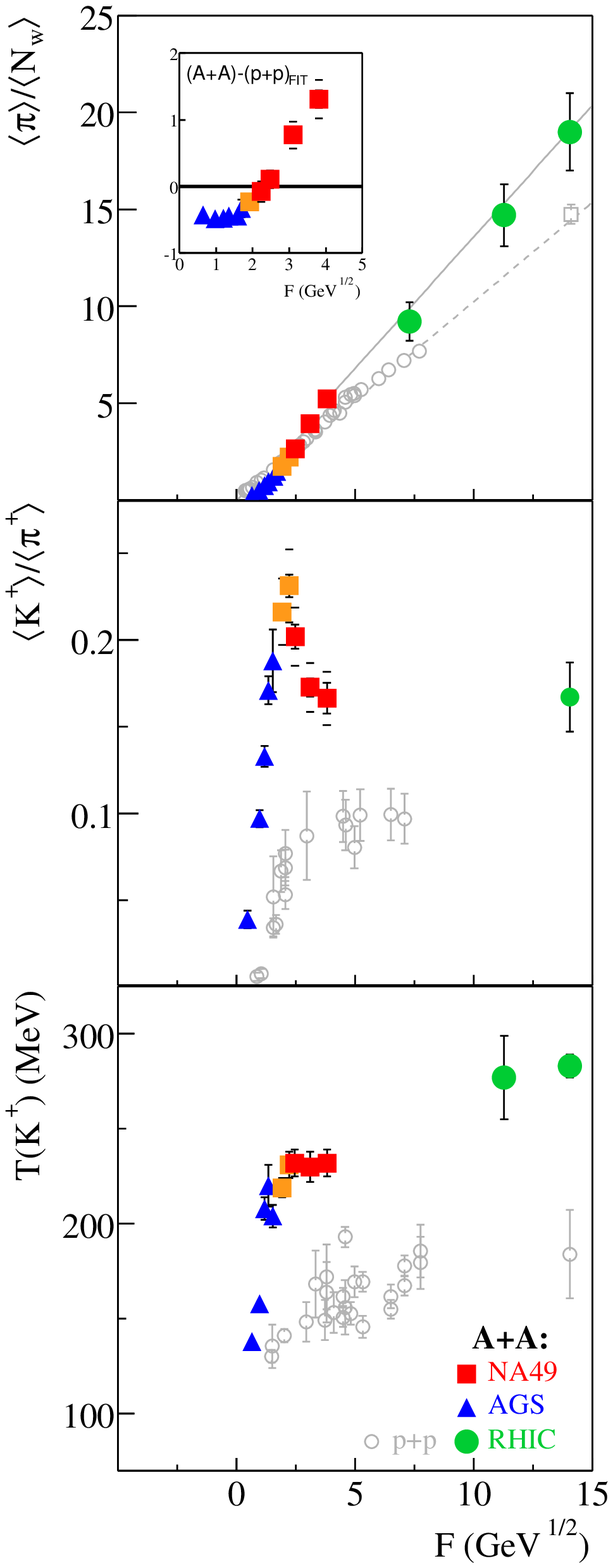,width=61mm} 
       \epsfig{file=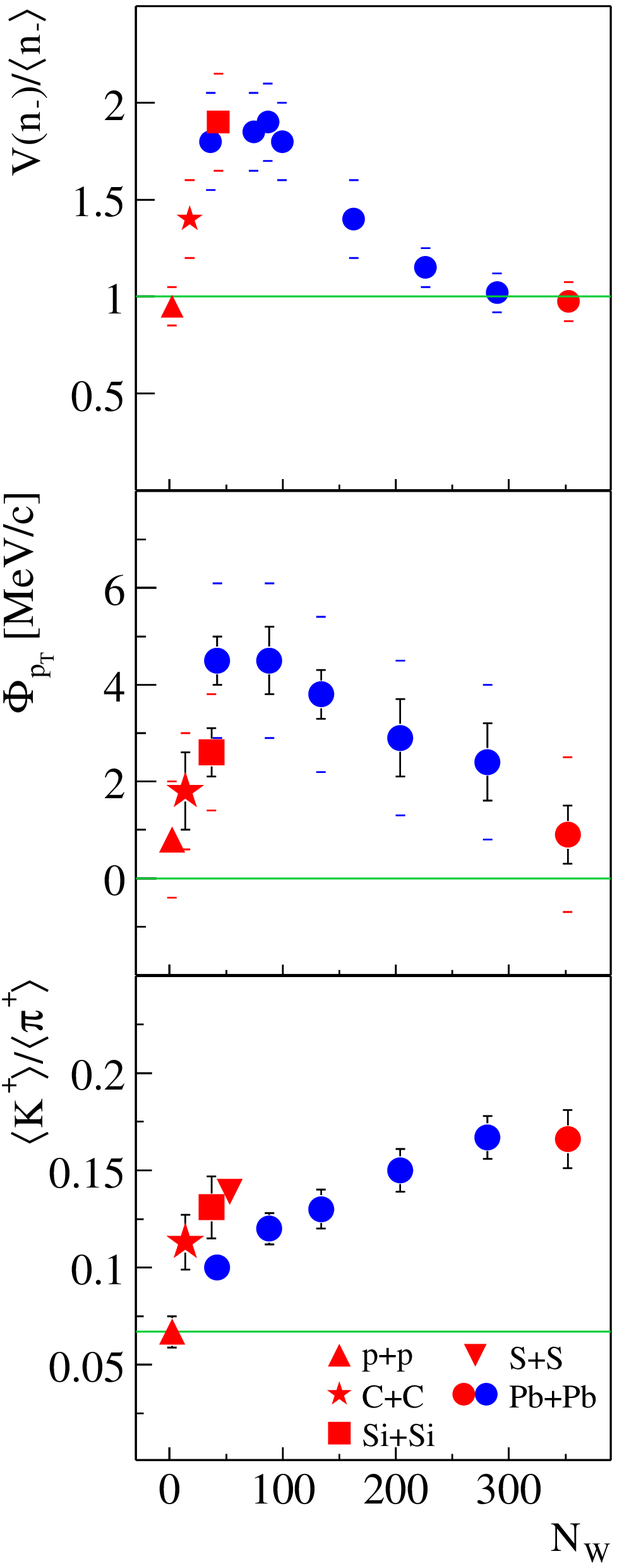,width=61mm}
         }
\end{center}
\vspace*{0.5cm}
\caption{
Left: The energy dependence of selected hadron
production properties measured in central Pb+Pb (Au+Au)
collisions (solid symbols) and p+p interactions (open symbols).
The changes in the SPS energy range (solid squares) suggest the
onset of the deconfinement phase transition.
Right: The system size dependence of the selected hadron production
properties at 158 $A$GeV.
Rapid changes are observed for small and intermediate mass
systems. In particular fluctuations in multiplicity and
transverse momentum reach a maximum at about $N_W \approx 70$. 
}
\label{sum}
\end{figure}

The energy dependence of hadron production in central Pb+Pb
collisions shows anomalies which are consistent with the
hypothesis that the onset of the deconfinement phase transition
is located at about 30 $A$GeV.
Selected results which illustrate this conclusion are
plotted in Fig.~\ref{sum} (left) using a common energy scale.

Large multiplicity and transverse momentum 
fluctuations are measured for intermediate mass systems 
($N_W \approx 70$) at 158 $A$GeV.
The corresponding results are summarized in  Fig.~\ref{sum} (right)
using a common system size scale.
For  comparison the  system 
size dependence of the
$\langle K^+ \rangle/\langle \pi^+ \rangle$ ratio measured 
by NA49 in the same reactions is also plotted.
The origin of the observed large fluctuations is unclear.


\begin{figure}
\begin{center}
\vspace*{-2cm}
\mbox{ \epsfig{file=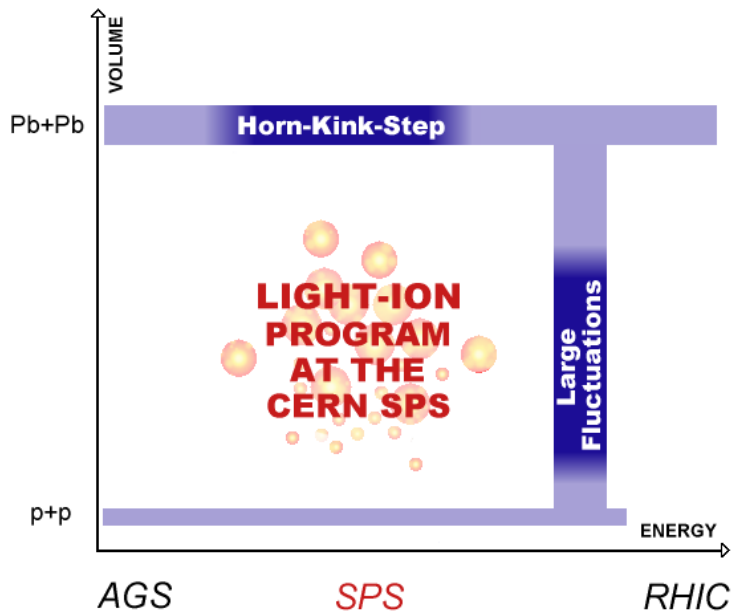,width=100mm} 
        }
\end{center}
\caption{
A schematic summary of the recent results at the CERN SPS.
The domain of  necessary future  studies is indicated.
}
\label{schem}
\end{figure}

A schematic graphical summary of the experimental situation
is shown in Fig.~\ref{schem}.
It is obvious that  the
turn on of the horn-kink-step like structures is located
in collisions of light or intermediate mass nuclei at the low
SPS energies.
It is not clear at all how the observed large fluctuations
at 158 $A$GeV will evolve with energy.  
Do they disappear at low energies, below the onset of deconfinement in
heavy ion collisions?
What are the fluctuations in the transition energy range? 

New measurements are needed in the SPS
energy range in order to answer these and other questions with the
final goal to understand the role played by 
the volume of  strongly interacting matter
in determining the onset of the deconfinement phase transition.
Recently an Expression of Interest \cite{eoi} for performing these, but
also other necessary measurements was submitted to the CERN SPS
Committee. 
This new exciting experimental study could start in three years 
from now.  

\vspace{1cm}
\noindent
Acknowledgements: This work was supported by the US Department of Energy
Grant DE-FG03-97ER41020/A000,
the Bundesministerium fur Bildung und Forschung, Germany, 
the Polish State Committee for Scientific Research (2 P03B 130 23, SPB/CERN/P-03/Dz 446/2002-2004, 2 P03B 04123), 
the Hungarian Scientific Research Foundation (T032648, T032293, T043514),
the Hungarian National Science Foundation, OTKA, (F034707),
the Polish-German Foundation, and the Korea Research Foundation Grant (KRF-2003-070-C00015).

\end{document}